\documentclass[twocolumn,aps,prl,showpacs,floatfix,superscriptaddress]{revtex4}
\usepackage{amsmath}
\usepackage{amsfonts}
\usepackage{amssymb}
\usepackage{bm}
\usepackage{color}
\usepackage{graphicx}

\begin{document}
\title{On the effects of vortex trapping on the velocity statistics of
  tracers and heavy particle in turbulent flows}

\author{J\'er\'emie Bec} \affiliation{CNRS UMR6202, Observatoire de la C\^ote
  d'Azur, BP4229, 06304 Nice Cedex 4, France.}

\author{Luca Biferale} \affiliation{Dipartimento di Fisica and INFN,
  Universit\`a di Roma ``Tor Vergata'', Via della Ricerca Scientifica
  1, 00133 Roma, Italy.}

\author{Massimo Cencini} \affiliation{INFM-CNR, SMC Dipartimento di
  Fisica, Universit\`a di Roma ``La Sapienza'',\\ Piazzale A.\ Moro, 2,
  00185 Roma, Italy.} \affiliation{ISC-CNR Via dei Taurini, 19, 00185
  Roma, Italy.}

\author{Alessandra S. Lanotte} \affiliation{CNR-ISAC, Str.\ Prov.\
Lecce-Monteroni, 73100 Lecce, Italy.}

\author{Federico Toschi} \affiliation{Istituto per le Applicazioni del
  Calcolo CNR, Viale del Policlinico 137, 00161 Roma, Italy.}
\affiliation{INFN, Sezione di Ferrara, Via G. Saragat 1, I-44100
  Ferrara, Italy.}

\date{\today}

\begin{abstract}
  The Lagrangian statistics of heavy particles and of fluid tracers
  transported by a fully developed turbulent flow are investigated by
  means of high resolution direct numerical simulations.  The
  Lagrangian velocity structure functions are measured in a time range
  spanning about three decades, from a tenth of the Kolmogorov
  timescale, $\tau_\eta$, up to a few large-scale eddy turnover time.
  Strong evidence is obtained that fluid tracers statistics are
  contaminated in the time range $\tau\in[1:10]\tau_\eta$ by a
  bottleneck effect due to vortex trapping. This effect is found to be
  significantly reduced for heavy particles which are expelled from
  vortices by inertia.  These findings help in clarifying the results
  of a recent study by \textit{H.\ Xu et al.}\/ [Phys.\ Rev.\ Lett.\
  {\bf 96} 024503 (2006)], where differences between experimental and
  numerical results on scaling properties of fluid tracers were
  reported.
\end{abstract}

                     
\maketitle 

\noindent Suspensions of dust, droplets, bubbles, and other
finite-size particles advected by incompressible turbulent flows are
commonly encountered in many natural phenomena and applied processes
ranging from cloud formation to industrial mixing
\cite{Falkovich_nature,industrial}. Understanding their statistical
properties is thus of primary importance. From a theoretical point of
view, the problem is more complicated than in the case of fluid
tracers, point-like particles with the same density of the carrier
fluid: inertia is responsible for the appearance of correlations
between the particle positions and the structure of the underlying
flow. It is well known indeed that heavy particles are expelled from
vortical structures, while light particles tend to concentrate in
their core. This results in the appearance of strong inhomogeneities
in the particle spatial distribution, often dubbed {\it preferential
concentration} \cite{Eaton,SE91}.  This phenomenon has recently
gathered much attention both from a theoretical
\cite{Falkovich-clustering,simo} and a numerical \cite{SE91,Collins2}
point of view.  Progresses in the statistical characterization of
particle aggregates have been achieved by studying particles evolving
in stochastic flows \cite{stuart,Falkovich-clustering,Bec-Gaw-Horvai}
and in two dimensional turbulent flows \cite{Boffetta}.  Concerning
single particle statistics, there has been considerable less attention
for heavy particles~\cite{JFM,wara}, in contrast with the numerous
studies devoted to tracers
\cite{borgas,boden,ott,pinton,MF,trapping,bodensf,yeung,yeung1}.  For
example, it is known that Lagrangian velocity structure functions of
tracers display high intermittent statistics~\cite{pinton,MF,bodensf},
while there are no results concerning heavy particles.\\ The aim of
this Letter is to compare the Lagrangian statistics of heavy particles
with that of fluid tracers evolving in the same turbulent flow, by
means of Direct Numerical Simulations (DNS). In particular, we shall
focus on the Lagrangian structure function of second order, important
for stochastic modelling of particle trajectories~\cite{P94,pdfmod},
and of higher orders, to account for intermittency effects.  For
Lagrangian studies, DNS offer the possibility to reach Reynolds
numbers comparable to experiments, with a full control of both
temporal and spatial properties of a very large number of tracers and
heavy particles.\\ We consider particles with a mass density $\rho_p$
much larger than the density $\rho_f$ of the carrier fluid. In this
limit, particles evolve according to ~\cite{maxey2}
\begin{equation}
\frac{d {\bm X}}{dt} = {\bm V} \; , \qquad \frac{d {\bm V}}{dt} =
-\frac{1}{\tau_s} \left[{\bm V}-{\bm u}({\bm X}(t),t) \right]\;.
\label{eq:1}
\end{equation}
${\bm X}(t)$ denotes the particle position, ${\bm V}(t)$ its velocity,
${\bm u}({\bm x},t)$ is the fluid velocity and $\tau_s =2 \rho_p a^2
/(9 \rho_f \nu)$ is the particle response time, where $a$ is the
radius and $\nu$ the fluid viscosity. The Stokes number, which
quantifies the degree of inertia, is defined as $St=\tau_s/\tau_\eta$,
where $\tau_\eta=(\nu/\epsilon)^{1/2}$ is the eddy turnover time
associated to the Kolmogorov scale and $\epsilon$ the average rate of
energy injection.  Equation (\ref{eq:1}) is derived in \cite{maxey2}
under the assumption of very dilute suspensions, where
particle-particle interactions (e.g.\ collisions) and hydrodynamic
coupling can be neglected. Tracers correspond to the limit $\tau_s\to
0$, i.e. to the evolution ${\rm d}\bm X/{\rm d} t= \bm u(\bm
X(t),t)$.\\ We performed a series of DNS of homogeneous and isotropic
turbulence on a cubic grid with resolutions up to $512^3$ (reaching a
Taylor micro-scale Reynolds number $R_\lambda \approx 180$) and
transporting millions of tracers and particles with $St \in
[0.16\!\!:\!\!1]$.  Particles are initially seeded homogeneously in
space with velocities equal to the local fluid velocity, already in a
stationary configuration. Then they evolve according to equation
(\ref{eq:1}) for about $2$ to $3$ large-scale eddy turnover times
before reaching a Lagrangian statistical steady state.  It is only
once the particle dynamics has completely relaxed that measurements
are started. Details on the simulation parameters can be found in
previous reports~\cite{JFM,JFM_JOT}.  For fluid tracers we also
present results obtained with resolution $1024^3$ and $R_{\lambda}
\approx 300$; these data are described in \cite{MF,trapping}.\\ The
Lagrangian Structure Functions (LSF)
\begin{equation} 
S^{(p)}(\tau) = \langle [ V(t+\tau)- V(t) ]^p \rangle\,, \label{eq:sf}
\end{equation}
measure the time variations of any component $V$ of the tracer or
particle velocity.  For time lags $\tau$ in the inertial range (i.e.\
when $\tau_\eta \ll \tau \ll T_L$, where $T_L$ is the integral
Lagrangian time scale), the LSF of tracers are expected to display
power law behaviors $S^{(p)}(\tau)\propto \tau^{\zeta(p)}$ (see
\cite{pinton} and references therein).  A dimensional estimate derived
from the Kolmogorov 1941 theory for Eulerian turbulence predicts
$S^{(p)}(\tau) \propto (\epsilon \tau)^{p/2}$.  Considerable
deviations from the non-intermittent scaling $\zeta(p)=p/2$ were
observed in both experiments \cite{pinton,bodensf} and DNS
\cite{MF,trapping}, with however some disagreement on the actual
values of the exponents.  For heavy particles there is not any
reference theory. As the effect of inertia on the velocity statistics
becomes less and less important when increasing $\tau/\tau_s$
\cite{Falkovich-clustering}, we can guess that particles recover the
statistical properties of tracers when $\tau_{\eta} \sim \tau_s \ll
\tau \ll T_L$.
%%%%%%%%%%%%%%%%%%%%%%%%%%%%%%%%%%%%%%%%%%%%%%%%%%%%%%%%%%%%%%%
\begin{figure}[!t]
\includegraphics[draft=false,scale=0.65]{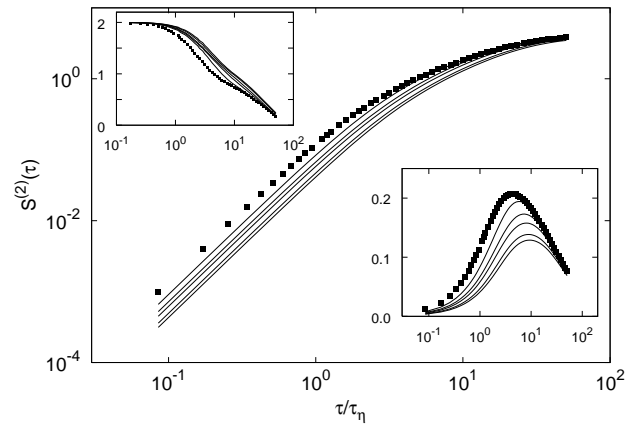}
\caption{$S^{(2)}(\tau)$ vs $\tau/\tau_{\eta}$ in log-log scale for
  tracers ($\blacksquare$) and heavy particles with $St=0.16$, $0.37$,
  $0.59$, $1.01$, and $1.34$ (solid lines from top to bottom).  Bottom
  inset: compensated plot $S^{(2)}(\tau)/\tau$ vs $\tau/\tau_{\eta}$,
  same symbols as in the body. Top inset: logarithmic derivative ${\rm
    d} \log(S^{(2)}(\tau))/{\rm d} \log (\tau)$ vs $\tau/\tau_{\eta}$,
  same symbols as in the body. For each Stokes number, averages in
  (\ref{eq:sf}) are performed over $N=5\times 10^5$ trajectories that
  last for about $200\tau_\eta$. All curves are obtained by averaging
  over the three components of the velocity vector to increase the
  statistics.}
\label{fig:1}
\end{figure}
%%%%%%%%%%%%%%%%%%%%%%%%%%%%%%%%%%%%%%%%%%%%%%%%%%%%%%%%%%%%%%%
%%%%%%%%%%%%%%%%%%%%%%%%%%%%%%%%%%%%%%%%%%%%%%%%%%%%%%%%%%%%%%%
\begin{figure}[!t]
\includegraphics[draft=false,scale=0.65]{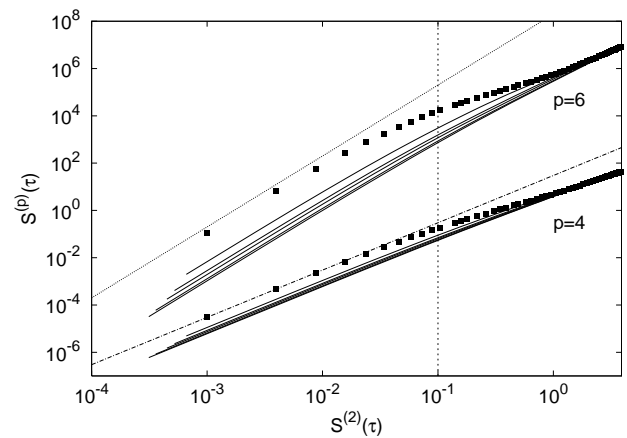}
\caption{Log-log plot of $S^{(p)}(\tau)$ vs $S^{(2)}(\tau)$ of order
  $p=4$ and $p=6$ for fluid tracers ($\blacksquare$) and heavy
  particles with $St=0.16$, $0.37$, $0.59$, $1.01$, and $1.34$ (solid
  lines from top to bottom). The two straight lines represent the
  dimensional non-intermittent scaling $\zeta(p)/\zeta(2) =
  p/2$. Notice the similarity of heavy particles for different Stokes
  and the marked difference of the tracers with respect to particles
  for values close to the viscous scale, $S^{(2)}(\tau_{\eta})\approx
  0.1$ {(vertical dotted line)}. }
\label{fig:2}
\end{figure}
%%%%%%%%%%%%%%%%%%%%%%%%%%%%%%%%%%%%%%%%%%%%%%%%%%%%%%%%%%%%%%%
\begin{figure*}[!t]
\includegraphics[draft=false,scale=0.65]{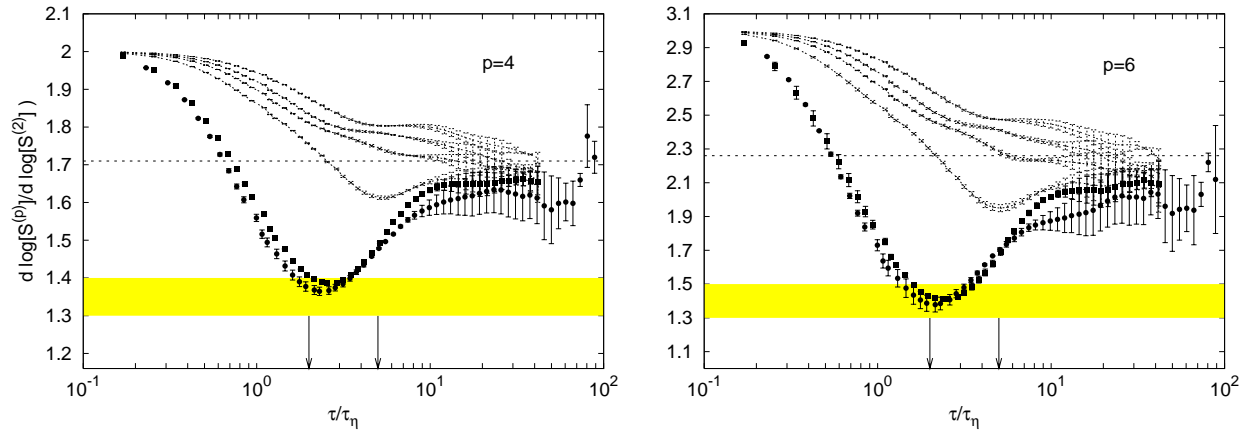}
\caption{(Color online) Logarithmic derivative of the ESS plots of
  Fig.~\ref{fig:2} vs $\tau/\tau_{\eta}$ for $p=4$ (left) and $p=6$
  (right). Data for fluid tracers at $Re_{\lambda}=180$ and
  $Re_{\lambda}=300$ are plotted as $\blacksquare$ and $\bullet$,
  respectively. The solid lines refer to heavy particles for
  $St=0.16,0.37,0.59,1.01$ (from bottom to top).  The dotted straight
  line marks the Lagrangian Multifractal prediction obtained for
  tracers in \cite{MF}
  ($\zeta(4)/\zeta(2)=1.71$,$\zeta(6)/\zeta(2)=2.26$.  The grey
  (yellow online) band corresponds to the experimental values measured
  for tracers in~\cite{bodensf} at comparable Reynolds numbers, and
  fitted in the range $\tau \in [2:5] \tau_{\eta}$ (shown with the
  arrows).  Errors have been estimated form the variations among the
  three velocity components. }
\label{fig:3}
\end{figure*}

\noindent 
We report in the sequel numerical measurements of the LSF of both
particles and tracers for time lags ranging from $\tau_\eta/10$ up to
$100\tau_\eta$ of the order of the integral time scale $T_L$.  Here we
anticipate the main findings. {Comparing the statistical properties of
  tracers and particles, we find further evidence that the formers are
  strongly affected by vortex trapping well above the Kolmogorov time
  scale, i.e. in the range $\tau \in [1\!\!:\!\!10]\tau_{\eta}$, while
  trapping becomes less and less important at increasing $St$.
  Trapping events spoil the scaling properties for time lags expected
  to be inside the inertial range of turbulence. This observation
  explains the disagreement, pointed out in \cite{bodensf}, existing
  between the scaling behavior of fluid tracers LSF measured in
  experiments \cite{boden,pinton} and DNS \cite{trapping,MF}.  We
  argue that this discrepancy stems from the fact that the
  experimental LSF scaling exponents are measured in the time range
  $[3\!:\!6]\tau_\eta$, exactly where trapping is effective. This
  leads to a more intermittent statistics than that measured in the
  inertial range \cite{trapping,MF}.}\\ Figure \ref{fig:1} summarizes
the results for the second order LSF at varying Stokes. It should be
noted that it is very difficult to identify a power-law scaling range
for any Stokes and any time lags. This is evident from the absence of
plateaux in the logarithmic derivatives of the LSF plotted in the top
inset. The apparent plateaux in the bottom inset, where we show the
compensated LSF normalized to the $S^{(2)}(\tau)/\tau$ at changing
Stokes, should thus not be considered as an evidence of a linear power
law.  { Still in the bottom inset, it is worth noticing the strong
  effects induced by inertia: already for the smallest Stokes number
  we considered ($St = 0.16$), we observe a clear departure from
  tracer behavior.\\ It is now interesting to look for non-trivial
  effects at higher-order moments. As customary in turbulence studies,
  assessing deviations from a simple dimensional scaling can be done
  by comparing all moments against a reference one, a procedure
  originally proposed for Eulerian structure functions and dubbed
  Extended Self Similarity (ESS)~\cite{ess}. This amounts to study the
  scaling behavior of the order $p$ LSF as a function of that of order
  $2$, used as a reference for Lagrangian statistics. This procedure
  is known to decrease finite-Reynolds effects and is frequently used
  for measuring scaling exponents in experiments and simulations.  The
  price to pay is that only relative scaling exponents
  $\zeta(p)/\zeta(2)$ can be measured.  Figure~\ref{fig:2} shows
  $S^{(p)}(\tau) $ as a function of $S^{(2)}(\tau)$ for $p=4$ and
  $p=6$ and various values of $St$.  Two observations can be done.
  First, the LSF of both inertial particles and tracers have an {\it
    inertial-range} scaling behavior that deviates significantly from
  the dimensional one $\zeta(p)/\zeta(2) = p/2$. Second, the
  differences between tracers and particles around $\tau \sim
  \tau_{\eta}$, i.e.\ for $S^{(2)}(\tau_{\eta}) \sim 0.1$, are now
  even more pronounced.}  To be more quantitative in Fig.~\ref{fig:3}
the local slopes ${\rm d} \log (S^{(p)}(\tau))/{\rm d} \log
(S^{(2)}(\tau))$ of all curves are shown.  Let us first focus on the
behavior well inside the {\it inertial range}. For $10 \tau_{\eta}<
\tau < T_L$, all the data sets display a tendency to converge, within
error bars, to the same scaling behavior as that of fluid tracers,
irrespectively of the value of the Stokes number. The large error bars
are due to unavoidable large-scale anisotropic fluctuations in the
statistics.  {In the same figure we also report the multifractal
  prediction for fluid tracers obtained in~\cite{MF} (see also
  \cite{borgas,boff}), by translating the well-known Eulerian
  multifractal model to the Lagrangian domain. Notice that the
  inertial-range behavior is compatible with the multifractal
  prediction also for heavy particles.\\ For time lags in the range
  $\tau_\eta\leq \tau \leq 10\tau_{\eta}$, the local scaling exponents
  reveal the presence a strong bottleneck effect, much more pronounced
  for tracers than particles. At these time scales, both for $p=4$ and
  $p=6$, fluid tracers have very intermittent scaling properties that
  are however smoothed out as soon as inertia is switched on.  This is
  due to the fact that tracers ($St=0$) may experience vortex trapping
  lasting for rather long times, while inertial particles (even for
  $St\ll 1$) are expelled from vortex filaments.  The values for the
  scaling exponents of the LSF given in the experimental study of
  fluid tracers in \cite{bodensf} at comparable Reynolds numbers are
  also represented in Fig.~\ref{fig:3} in the shaded band.  In
  Ref.~\cite{bodensf}, the scaling exponents are measured in the range
  $\tau \in [2:5] \tau_{\eta}$ (see arrows in the figure), where
  trapping into vortex filaments gives the dominant contribution
  leading to a more intermittent statistics.  For these time lags, the
  values of the scaling exponents reported in \cite{bodensf} are in
  good agreement with DNS results.  However, these exponents
  substantially differ from the inertial-range values of the
  logarithmic derivative that we observe at $\tau > 10 \tau_{\eta}$.}
\\ \noindent It is worth stressing that the statistical signature of
particle trapping into vortex filaments has already been the subject
of experimental and numerical studies of fully developed turbulent
flows.  In particular, it was shown that such events play a crucial
role in determining the intense fluctuations of tracer
acceleration~\cite{boden,pinton,MF,bif1}.  This effect has previously
been highlighted by filtering out the intense vortical events in the
LSF (see Fig.~3 of \cite{trapping} and compare it with
Figs.~\ref{fig:1} and \ref{fig:2}). Such a filtering is here obtained
dynamically by switching on inertia, whose major landmark at small
values of the Stokes number is to expel heavy particles from vortex
filaments.\\
{
  \noindent In conclusion, we have shown that for large-enough time
  lags ($\tau > 10 \tau_{\eta}$) the scaling properties of heavy
  particles with $St \lesssim 1$ tend to be almost independent of
  $St$, within error bars. The relative scaling exponents of LSF for
  the investigated Stokes numbers are found in the range
  $\zeta(4)/\zeta(2) \in [1.55\!:\!1.75] $ for the fourth order and
  $\zeta(6)/\zeta(2) \in [1.8\!:\!2.4] $ for the sixth one.
  Investigations at larger $Re_\lambda$ are needed to confirm the
  robustness of the universality enjoyed by heavy particles for large
  time lags.}  This observation can be explained by the fact that the
effect of inertia on the particle velocity should disappear
progressively when increasing the
scale~\cite{Falkovich-clustering}. For time lags well inside the
inertial range and much larger than the response time $\tau_s$, the
delay of the particles on the fluid motion becomes negligible and a
``tracer-like'' physics is recovered \footnote{A study on
  white-in-time rough stochastic flows shows that heavy particles, at
  large times, display a Richardson diffusion, akin to that observed
  for tracers advected by the same flow -- J.\ Bec, M.\ Cencini and
  R.\ Hillerbrand, in preparation.}.  For the case of fluid tracers,
it is important to be very careful in assessing scaling properties
because trapping into vortex filaments may spoil the {\it inertial
  range} scaling behavior for time lags up to $\tau \sim 10
\tau_{\eta}$ and even more. Such trapping effects are less effective
for heavy particles due to their dynamical properties that bring them
outside of strong vortex filaments.  These results are important for
stochastic modelling of both tracers and heavy particles, where the
statistical properties of velocity and acceleration along the
trajectories are the main ingredients for developing appropriate
models. \\

\noindent We acknowledge useful discussions with E. Bodenschatz,
G. Boffetta, A. Celani. This work has been partially supported by the
EU under contract HPRN-CT-2002-00300 and the Galileo programme on
Lagrangian Turbulence. Numerical simulations have been performed at
CINECA (Italy) and IDRIS (France) under the HPC-Europa programme,
contract number RII3-CT-2003-506079.  We thank also the ``Centro
Ricerche e Studi Enrico Fermi'' and N.~Tantalo for technical
support. Raw, unprocessed data analyzed in this study are freely
available from the iCFDdatabase web site (http://cfd.cineca.it).

\end{document}